\documentclass[]{AO4ELT}  

\usepackage{microtype}
\usepackage{biblatex}
\usepackage{amsmath,amsfonts,amssymb}
\usepackage{graphicx}
\usepackage{pst-all} 
\usepackage[colorlinks=true, allcolors=blue]{hyperref}
\usepackage{mathtools}
\DeclarePairedDelimiter\bra{\langle}{\rvert}
\DeclarePairedDelimiter\ket{\lvert}{\rangle}
\DeclarePairedDelimiterX\braket[2]{\langle}{\rangle}{#1\,\delimsize\vert\,\mathopen{}#2}

\makeatletter         
\def\@maketitle{
\includegraphics[width = 170mm]{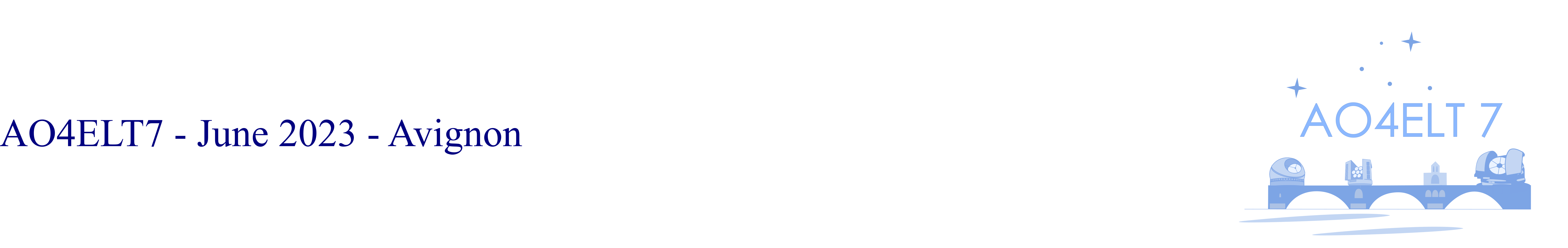}\\[8ex]
\begin{center}
{\Huge \bfseries \sffamily \@title }\\[4ex] 
{\Large  \@author}\\[4ex] 
\@date
\end{center}}

\title{Reaching the fundamental sensitivity limit of wavefront sensing on arbitrary apertures with the Phase Induced Amplitude Apodized Zernike Wavefront Sensor (PIAA-ZWFS).}

\author[a]{Sebastiaan Y. Haffert}
\author[a]{Jared Males}
\author[a,b,c,d]{Olivier Guyon}
\affil[a]{University of Arizona, Steward Observatory, Tucson, Arizona, United States}
\affil[b]{National Astronomical Observatory of Japan, Subaru Telescope, National Institutes of Natural Sciences, Hilo, HI 96720, USA}
\affil[c]{Wyant College of Optical Science, University of Arizona, 1630 E University Blvd, Tucson, AZ 85719, United States}
\affil[d]{Astrobiology Center, National Institutes of Natural Sciences, 2-21-1 Osawa, Mitaka, Tokyo, JAPAN}

\authorinfo{Further author information: (Send correspondence to S.Y.H.)\\S.Y.H.: E-mail: shaffert@arizona.edu}

\begin{document} 
\maketitle
\begin{abstract}
In the last two decades many people have been searching for the optimal wavefront sensor as it can boost the performance of high-contrast imagining by orders of magnitude on the ELTs. According classical information theory, the optimal sensitivity of a wavefront sensor is 1/2 radian rms per photon. We show that classical limit is also the quantum metrology limit for starlight, which means that 1/2 radian rms per photon is really the limit. This proceeding introduces the Phase Induced Amplitude Apodized Zernike Wavefront sensor. The PIAA-ZWFS modifies a standard ZWFS with a set of aspheric lenses to increase its sensitivity. The optimized system reaches the fundamental limit for all spatial frequencies $>$1.7 cycles/pupil and is very close to the limit for the spatial frequencies $<1.7$ cycles/pupil. The PIAA-ZWFS can be seamlessly integrated with the PIAA-CMC coronagraphy. This makes the PIAA-ZWFS an ideal candidate as wavefront sensor for high-contrast imaging.
\end{abstract}


\section{INTRODUCTION}
\label{sec:intro}
Adaptive optics (AO) has evolved tremendously over the past few decades and is now a crucial part of any large ground-based telescope. A fundamental part of any AO system is the wavefront sensor. There are many wavefront sensors each using a different way to encode the wavefront into an intensity measurement. There are many competing requirements that influence the choice of wavefront sensor. However, the most important part is the WFS sensitivity. The sensitivity ultimately determines the limit on guide star faintness and the AO loop speed. Improving the WFS sensitivity would allow us to improve both. High-contrast imaging is one of the most demanding applications of AO and an optimal wavefront sensor would boost the performance of high-contrast imagining by orders of magnitude on the ELTs \cite{guyon2005limits}.

The first generation of AO systems used curvature wavefront sensors (CWS) \cite{roddier1988curvature, rigaut1998performance} and Shack-Hartmann wavefront sensors (SHWFS) \cite{shack1971production,beuzit1997adaptive}. These provided diffraction-limited performance at wavelengths upwards of 2 um. The AO systems driven by SHWFSs were the very first to directly image faint sub-stellar companions \cite{chauvin2004giant}.  However, it was also shown that the SHWFS has relatively poor sensitivity for low-order modes \cite{guyon2005limits}. A major step was made by the introduction of the pyramid wavefront sensor (PWFS) \cite{ragazzoni1996pupil}. The PWFS focuses a beam onto the tip of a pyramidal prism. Each face of the pyramid sends the light into a different direction which results in four pupils if there are four faces. The PWFS by itself has a limited dynamic range. If the Point Spread Function (PSF) moves of the tip of the pyramid all the light is concentrated into one pupil saturating the reconstruction. This is resolved by modulating the PSF on the tip of the pyramid. The modulation makes the PSF bigger which increases the dynamic range at the cost of sensitivity \cite{ragazzoni1996pupil,verinaud2004nature}. Many AO systems now use the PWFS \cite{pinna2016soul,lozi2019visible,males2022magao} and, it is also baselined for all new high-contrast imaging systems \cite{kasper2021pcs, males2022conceptual,haffert2022visible,fitzgerald2022planetary}, and will be part of upgrades of current AO systems \cite{boccaletti2022upgrading,perera2022gpi}.

While the PWFS is a major step forward, it is not the most sensitive wavefront sensor. The Zernike wavefront sensor (ZWFS) is an example of a WFS with higher sensitivity. The ZWFS is a common-path point diffraction interferometer. It focuses the beam onto a phase mask with a circular phase dimple that has a similar size to the core of an Airy pattern. The high-spatial frequency sensitivity can increased by changing the size of the dimple at the cost of a loss of sensitivity at low-spatial frequencies \cite{chambouleyron2021variation}. It is also possible to optimize the full focal plane mask instead of only the focal plane dot. The mask can be optimized by allowing either a purely linear or even non-linear wavefront responses \cite{chambouleyron2022optimizing, landman2022joint}. The fully optimized masks were able to achieve an even higher sensitivity than the optimized ZWFS for spatial frequencies above 5 cycles per pupil at the cost of lower sensitivity for the low-order modes which are critical for high-contrast imaging. 

The examples covered here show that it is possible to improve wavefront sensors and keep achieving higher sensitivity by modifying the sensor design.  But what is actually the fundamental sensitivity limit of a wavefront sensor? Is there an upper limit? Classical information theory gives us an hint at the answer. The variance of any unbiased wavefront estimator is bound by the Cramer-Rao bound (CRB). The CRB tells us that the variance is bounded by the inverse of the Fisher Information (FI). This analysis derives a sensitivity bound of $1/2$ radian rms per photon for a discrete set of orthogonal modes \cite{paterson2008towards}. A different approach based on linear transfer functions arrives at the same answer. This approach comes to a result that it is possible to achieve $1/2$ radian rms per photon for any spatial frequency \cite{fauvarque2017general, chambouleyron2022optimizing}. However, this can not be true because piston from a single aperture is impossible to measure. There has to be a transition between no sensitivity and optimal sensitivity as a function of spatial frequency. Understanding this behavior is important because atmospheric turbulence creates wavefront errors at all spatial scales due to its power-law behavior. There is no finite discrete set of wavefront modes that fully describes turbulence. Also, there are several astronomical instruments that are trying to image faint companions at sub-Rayleigh distances. This begs the question; what is actually the fundamental limit? And is it possible to develop a sensor that approaches this limit.
 
In this work we will discuss the fundamental sensitivity limit for fully coherent source from the perspective of quantum optics. In Section 3 we propose a wavefront sensing architecture that can achieve this bound for arbitrary aperture shapes. 
Section 4 is a numerical evaluation of the performance.

\section{FUNDAMENTAL LIMIT}


\subsection{Cramer-Rao Bound}
The sensitivity of a wavefront sensor can be derived by using the Cramer-Rao bound (CRB). The CRB tells us that the variance of an unbiased parameter estimator, $\hat{\theta}$, after $N$ measurements is bounded by the inverse of the Fisher Information $F(\theta)$,
\begin{equation}
    Var[\hat{\theta}] \geq \frac{1}{N F(\theta)}.
\end{equation}
Quantum information theory tells us that there are two Fisher information metrics. The first is the classical Fisher information metric and the second is the Quantum Fisher information metric. The Fisher information metric of a particular measurement is bounded by the classical Fisher information, which in turn is bounded again by the Quantum Fisher information,
\begin{equation}
    F[\theta] \leq F_{C}[\theta] \leq F_{Q}[\theta].
\end{equation}
Here $F_{C}[\theta]$ is the classical Fisher Information which is defined as,
\begin{equation}
    F_{C}[\theta] = \mathrm{E}\left[\left(\frac{\partial}{\partial \theta} \log{f(\vec{r};\theta)}\right)^2\;\middle|\; \theta\right] =\int_{A} \left(\frac{\partial}{\partial \theta} \log{f(\vec{r};\theta)}\right)^2 f(\vec{r};\theta) \mathrm{d}\vec{r}.
\end{equation}
The measurement probability distribution is defined as $f(\vec{r};\theta)$ with $\vec{r}$ the position vector across the pupil. If the probability distribution is well behave it may be described as,
\begin{equation}
    F_{C}[\theta] = \int_{A} \frac{\left(\partial_\theta f(\vec{r};\theta)\right)^2}{f(\vec{r};\theta)} \mathrm{d}\vec{r}.
\end{equation}

Let's now derive the Fisher information for the phase under the assumption of photon noise. We assume that our integration time is short enough that we only need to consider a single photon. The output intensity distribution $I(\vec{r}; \theta)$ is then the probability distribution of where the photon will land. This means that the expectation operator will integrate over all spatial positions,
\begin{equation}
    F_{C}[\phi] = \int_{A} \frac{\left(\partial_\phi I(\vec{r};\phi)\right)^2}{I(\vec{r};\phi)} \mathrm{d}\vec{r}.
\end{equation}
Suppose that the spatial position is discretized, which naturally happens when a detector makes a measurement. The intensity can then be described by a vector, $I(\vec{r_i})=\vec{I}_i$. This also allows us to replace the integral by a summation,
\begin{equation}
    F_{C}[\phi] = \sum_i \frac{\left(\partial_\phi I_i\right)^2}{I_i}.
\end{equation}
A shorter way to write this is through the norm operator,
\begin{equation}
\label{eq:discrete_fi}
    F_{C}[\phi] = \left \| \frac{\partial_\phi \vec{I}}{\sqrt{\vec{I}}} \right \|^2.
\end{equation}
The norm operator for a vector is defined as $\left \| x \right \|=\sqrt{x\cdot x}$. Equation \ref{eq:discrete_fi} shows that the typical definition of a wavefront sensors sensitivity \cite{chambouleyron2022modeling} is actually the classical Fisher information. With this we arrive at the reconstructed phase coefficient variance as,
\begin{equation}
    Var[\hat{\phi}] \geq \frac{1}{N F_C(\phi)}.
\end{equation}
Here, $N$ is the number of repeated measurements or in other words the number of photons. A bound of $F_C(\phi)\leq4$ was found for a set of discrete modes that were mixed through an unitary optical system \cite{paterson2008towards}. Therefore,
\begin{equation}
    Var[\hat{\phi}] \geq \frac{1}{4N}.
\end{equation}

\subsection{Quantum Fisher Information}
The CFI is bounded by the Quantum Fisher Information. Therefore, it might be possible to make more sensitive measurements than classical theory tells us. Here, we will derive the QFI bound for starlight. Under the weak-source approximation for thermal sources, which is valid for starlight, the state of the field over a single temporal coherence interval is in the vaccuum state with probability $1-\epsilon$ or sometimes in a single-photon excitation state with probability $\epsilon \ll 1$ \cite{tsang2016quantum}. The density matrix of the system can then be described by,
\begin{equation}
    \rho_{T} = (1-\epsilon) \ket{0}\bra{0} + \epsilon \rho + \mathcal{O}(\epsilon^2). 
\end{equation}
Here $\rho_{T}$ is the density matrix of all states and $\rho$ is the density matrix of just the single-photon excitation state. Higher-order $\epsilon$ terms can be neglected for starlight \cite{tsang2016quantum}. The single-photon excitation state is,
\begin{equation}
    \rho = \ket{\psi}\bra{\psi},
\end{equation}
where $\bra{\psi}$ is single-photon state of the electric field at the entrance of our optical system. The state is normalized such that $\braket{\psi}{\psi}=1$. The density state is completely determined by a single wave function, which means it is a pure state. This can be easily verified because for a pure state $\rho^2 = \rho$. Working this out for the proposed density state leads to
\begin{equation}
    \rho^2 = \rho \rho = \ket{\psi}\braket{\psi}{\psi}\bra{\psi} =\ket{\psi}\bra{\psi} = \rho .
\end{equation}
From this it is clear that the proposed density matrix represents a pure quantum state. The Quantum Fisher information matrix (QFIM) for a pure quantum state is defined as \cite{braunstein1994statistical, pezze2017optimal},
\begin{equation}
    [F_{Q}]_{n,m} = 4\Re{\braket{\partial_{\alpha_n} \psi}{\partial_{\alpha_m} \psi}} - 4\braket{\partial_{\alpha_n} \psi}{ \psi} \braket{\partial_{\alpha_m} \psi}{ \psi}.
\end{equation}
Here the partial derivatives are with respect to the to be estimated parameters $\alpha_n$ and $\alpha_m$. The QFIM can be evaluated if a representation of the quantum state is chosen. The representation of the electric field in the pupil is,
\begin{equation}
    \ket{\psi} = \int_{\infty}^{\infty}\int_{\infty}^{\infty} \ket{x,y} \psi(x,y) \mathrm{d}x\mathrm{d}y.
\end{equation}
Here $\psi(x,y)$ can be treated in the same way as a classical electric field. Which means that an aberrated wavefront is given by,
\begin{equation}
    \psi = Pe^{i\sum_n a_n \phi_n}.
\end{equation}
The coefficients $a_n$ are the modal coefficients of the wavefront in some basis $\{\phi_n\}$ and $P$ is the pupil function which is 1 within the pupil and 0 outside. The first term of the QFIM is
\begin{equation}
    \braket{\partial_{\alpha_n} \psi}{\partial_{\alpha_m} \psi} = \int_A -i\phi_n P e^{-i\sum_n a_n \phi_n} i\phi_m P e^{i\sum_n a_n \phi_n}\mathrm{d}A = \int_A \phi_n \phi_m P \mathrm{d}A.
\end{equation}
Now suppose that our basis is defined to only exists on the pupil, $\phi_n P = \phi_n$, then the first term of the QFIM just measures the correlation between different modes,
\begin{equation}
    \braket{\partial_{\alpha_n} \psi}{\partial_{\alpha_m} \psi} = \int_A \phi_n \phi_m \mathrm{d}A.
\end{equation}
The second term of the QFIM measure the correlation between the partial derivative and the state itself. In other words it measures the similarity of a change in the state to the state itself.
\begin{equation}
    \braket{\partial_{\alpha_n} \psi}{\psi} = \int_A -i\phi_n P e^{-i\sum_n a_n \phi_n} P e^{i\sum_n a_n \phi_n}\mathrm{d}A = \int_A \phi_n P^2 \mathrm{d}A = \int_A \phi_n \mathrm{d}A.
\end{equation}
Here we can see that the second term measures the similarity between the phase mode and the piston term. So now suppose we have a mode basis that is orthogonal on the support of the pupil and the first mode is piston. Then $\braket{\partial_{\alpha_n} \psi}{\partial_{\alpha_m} \psi}=\delta_{nm}$ and $\braket{\partial_{\alpha_n} \psi}{\psi}=\delta_n$. Therefore, the QFIM is,
\begin{equation}
    [F_{Q}]_{n,m} = \left\{\begin{matrix}
0 & n=m=0\\ 
0 & n \neq m\\ 
4 & n=m
\end{matrix}\right.
\end{equation}
Here we see that $F_Q=4$ for all modes except piston. This matches with the earlier derived classical bound which shows that for starlight the classical Fisher Information bound is also the Quantum Fisher Information bound. That means that the previously derived bound of CFI of 4 is also the fundamental limit. If a measurement saturates the CFI, it will also saturate the QFI and it will therefore be the most sensitive wavefront sensor for starlight.


\section{PIAA-ZWFS}
Any interferometer reaches its maximum signal-to-noise ratio when the two interfering beams have an equal amplitude and opposite phase resulting in perfect destructive interference. The ZWFS is an interferometer and fundamentally already a very sensitive wavefront sensor. However, the response of the focal plane mask is an Airy pattern profile which is different from the incoming top hat profile. Previous work found that optimizing the full focal plane mask lead to a reference electric field that tried to mimic this top hat profile \cite{chambouleyron2022optimizing, landman2022joint}. 

\begin{figure}
    \begin{center}
        \includegraphics[width=\textwidth]{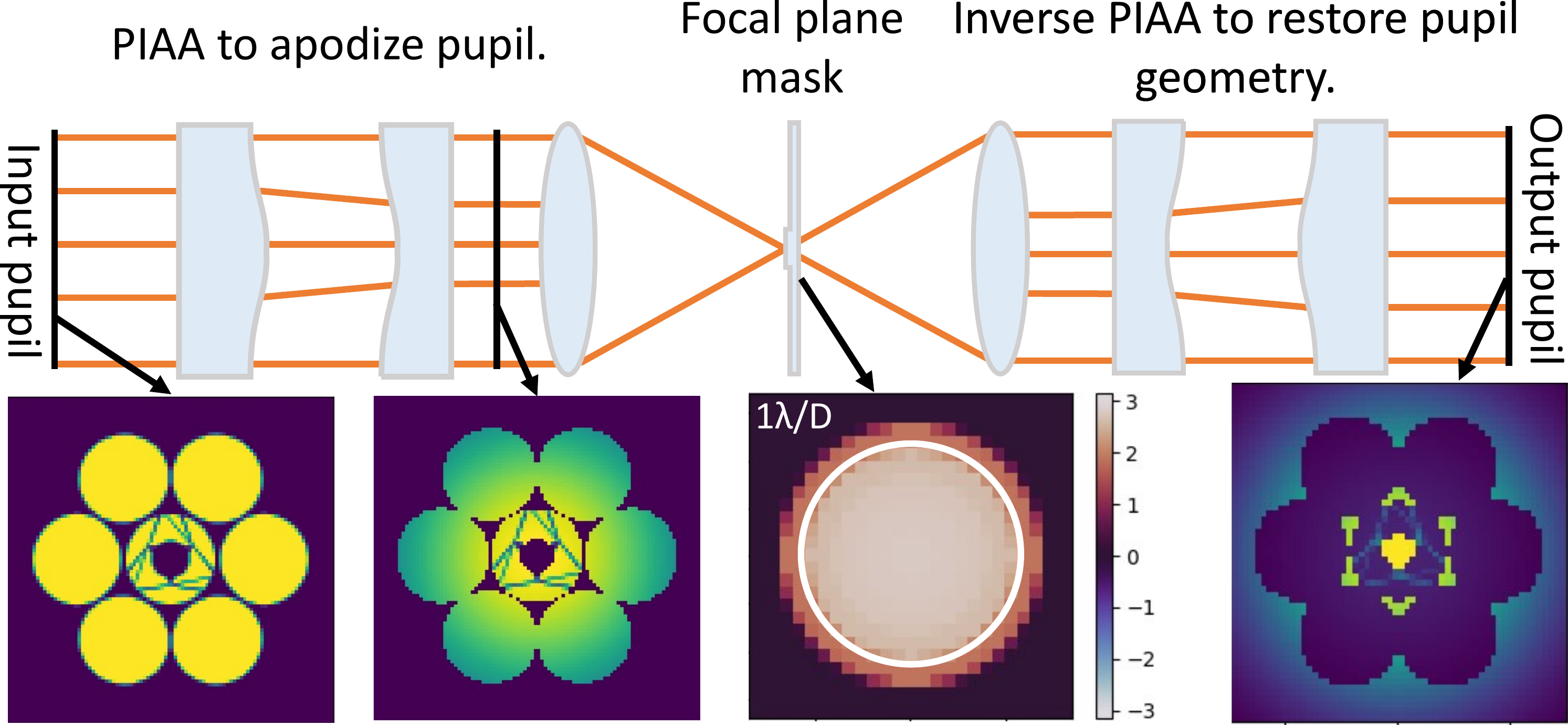}
	\end{center}
    \caption{The Phase Induced Amplitude Apodized Zernike Wavefront Sensor (PIAA-ZWFS) architecture. The incoming beam is apodized losslessly by a set of aspheric optics to match the intrinsic response of the focal plane mask. The partially phase-shifting mask will then create the optimal reference beam for interferometry. A second set of PIAA lenses in reverse are used to restore the original pupil geometry.}
    \label{fig:piaacmc} 
\end{figure}

An easier way to achieve the mode matching is to apodize the incoming aperture. Apodizing the aperture with just a grayscale mask would defeat the purpose because while each photon would carry more information we also lose photons from the apodization. Therefore, we propose to use a set of aspheric optics that reshape the distribution of light. This sensor is named the Phase Induced Amplitude Apodized Zernike Wavefront Sensor (PIAA-ZWFS). This argument of mode matching is exactly the same argument that was used in the development of the PIAA coronagraph \cite{guyon2010high}. The PIAA coronagraphs use the aspheric lenses to achieve a deeper coronagraph residual.

The input aperture of the optimization process is the GMT aperture. This is a challenging aperture because of its segmented nature and non-circular shape. We evaluate the performance of the different variations of the ZWFS for spatial frequencies from 0 to 15 cycles/pupil. This is a quite a small range compare to the expected control region of GMagAO-X \cite{males2022conceptual}. However, the results will show that the sensitivity is constant from about 5 cycles/pupil. The reference ZWFS uses a mask with a 2 $\lambda/D$ diameter which was suggested to be more sensitive \cite{chambouleyron2021variation}. It would only be a fair comparison if the mask size of the ZWFS is also chosen for sensitivity. The second design uses an optimized focal plane mask without the PIAA lens apodization. This mask is optimized by considered the sensitivity of spatial frequencies between 1 and 15 cycles/pupil. Lower spatial frequencies should not be included because those are fundamentally very difficult to sense. The PIAA-ZWFS focal plane mask is optimized together with the apodization profile.

\begin{figure}
    \begin{center}
        \includegraphics[width=\textwidth]{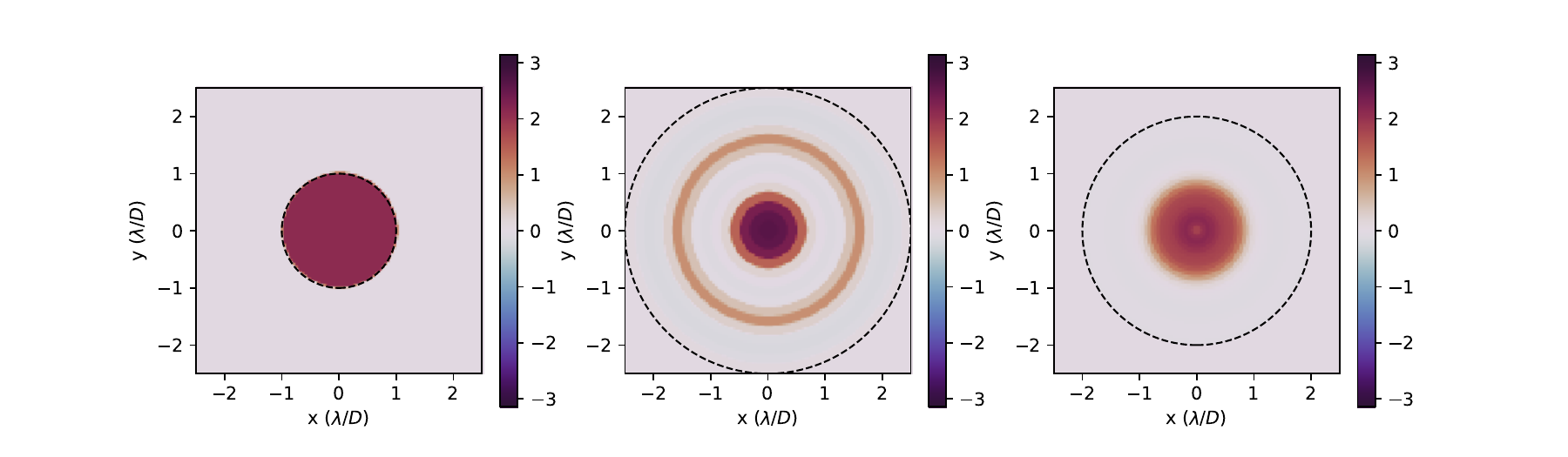}
	\end{center}
    \caption{The Phase Induced Amplitude Apodized Zernike Wavefront Sensor (PIAA-ZWFS) architecture. The incoming beam is apodized losslessly by a set of aspheric optics to match the intrinsic response of the focal plane mask. The partially phase-shifting mask will then create the optimal reference beam for interferometry. A second set of PIAA lenses in reverse are used to restore the original pupil geometry.}
    \label{fig:fpm} 
\end{figure}

Each of these three masks can be seen in Figure \ref{fig:fpm}. The optimized results show that the focal plane mask without apodization starts to mimic the PSF to create a tophat-like reference beam. Earlier work also found this result \cite{chambouleyron2022optimizing,landman2022joint}. The PIAA-ZWFS mask converges to a compact solution ($\sim$ 1 lambda/D diameter) that is not purely $\pi/2$.

\begin{figure}
    \begin{center}
        \includegraphics[width=\textwidth]{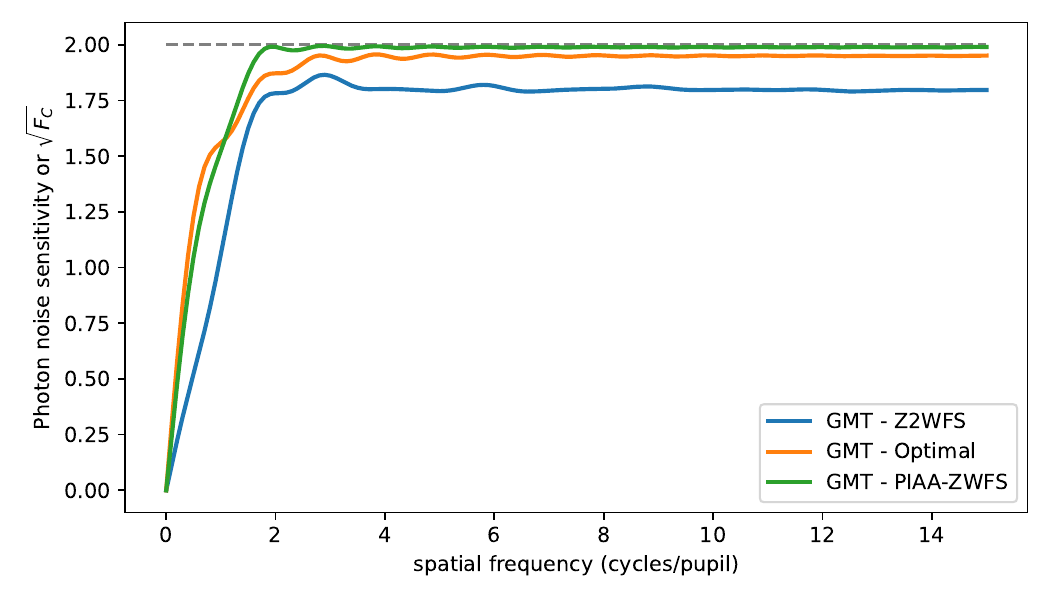}
	\end{center}
    \caption{The Phase Induced Amplitude Apodized Zernike Wavefront Sensor (PIAA-ZWFS) architecture. The incoming beam is apodized losslessly by a set of aspheric optics to match the intrinsic response of the focal plane mask. The partially phase-shifting mask will then create the optimal reference beam for interferometry. A second set of PIAA lenses in reverse are used to restore the original pupil geometry.}
    \label{fig:sensitivity} 
\end{figure}

The sensitivity as a function of spatial frequency can be found in Figure \ref{fig:sensitivity}. The PIAA-ZWFS reaches a sensitivity of 2 for all spatial frequencies above 1.7 cycles/pupil. The sensor reaches a sensitivity of 1.65 at 1 cycles/pupil. We see a decay of the sensitivity as we go to lower spatial frequencies. This is expected because the sensitivity will go down when the mode starts to become more similar to piston mode (as discussed in Section 2.2). This means that the PIAA-ZWFS achieves the fundamental limit $>$1.7 cycles per pupil and is very close for the limit for the other frequencies. The PIAA-ZWFS is almost twice as sensitive as the classic ZWFS at 1 cycle/pupil and 15\% more sensitive for all other modes. The optimal phase mask sensitivity saturates around a sensitivity of 1.9. The PIAA-ZWFS mostly gains over the optimal mask between 1.5 and 2.5 cycles/pupil.

\section{CONCLUSION}
We have shown that the classical sensitivity limit of wavefront sensing is the same as the quantum metrology limit. Both methods show that the maximum Fisher Information is $F_C=F_Q=4$. We have also shown that what literature has been calling the wavefront sensitivity is nothing else than the square root of the Fisher Information. The analysis shows that any wavefront sensor that saturates the classic Fisher information will be the most sensitive wavefront sensor that you can make. 

The ZWFS is a wavefront sensor with very high sensitivity. The focal plane mask creates the reference beam for the interfere. The loss of sensitivity of the ZWFS as compared to the fundamental limit is caused by a mode mismatch between the reference beam and the pupil distribution (tophat). We have shown that this can be solved by using a set of aspheric lenses that losslessly redistribute the light to create an apodized profile. We optimized both the pupil apodization and focal plane mask together to find the optimal system. The optimized system reaches the fundamental limit for all spatial frequencies $>$1.7 cycles/pupil and is very close to the limit for the spatial frequencies $<1.7$ cycles/pupil.

The extreme AO system MagAO-X already has a set of PIAA lenses for the MagAO-X PIAACMC. We will investigate the development of a PIAA-ZWFS for MagAO-X using the already manufactured PIAACMC lense.

\acknowledgments      
 Support for this work was provided by NASA through the NASA Hubble Fellowship grant \#HST-HF2-51436.001-A awarded by the Space Telescope Science Institute, which is operated by the Association of Universities for Research in Astronomy, Incorporated, under NASA contract NAS5-26555 and by the generous support of the Heising-Simons Foundation.

\printbibliography
\end{document}